\documentclass[aps,preprint,prl,amsmath,superscriptaddress,floatfix,showpacs,showkeys]{revtex4-1}
\usepackage{graphicx}% Includibe figure files
\usepackage{dcolumn}% Align table columns on decimal point
\usepackage{bm}% bold math
\usepackage{color}
\usepackage{amsmath}
\usepackage[normalem]{ulem}

\newcommand{\be}{\begin{equation}}
\newcommand{\ee}{\end{equation}}
\newcommand{\pa}[1]{\left(#1\right)}
\begin{document}
\preprint{IIPDM-2019}
\title{Detecting Dark Matter with Neutron Star Spectroscopy}

%\thanks{}%

\author{Daniel A. Camargo, Farinaldo S. Queiroz, Riccardo Sturani}

\affiliation{International Institute of Physics, Universidade Federal do Rio Grande do Norte, Campus Universit\'ario, Lagoa Nova, Natal-RN 59078-970, Brazil\\}%

\date{\today}% It is always \today, today,
\begin{abstract}

The presence of dark matter has been ascertained through a wealth of astrophysical and cosmological phenomena and its nature is a central puzzle in modern science. Elementary particles stand as the most compelling explanation. They have been intensively searched for at underground laboratories looking for an energy recoil signal and at telescopes sifting for excess events in gamma-ray or cosmic-ray observations. In this work, we investigate a detection method based on spectroscopy measurements of neutron stars. We outline the luminosity and age of neutrons stars whose dark matter scattering off neutrons can heat neutron stars up to a measurable level. We show that in this case neutron star spectroscopy could constitute the best probe for dark matter particles over a wide masses and interactions strength.  

\end{abstract}

% \pacs{Valid PACS appear here}% PACS, the Physics and Astronomy
\maketitle

\section{Introduction}
%\newline \indent 
We have collected several solid evidence for the presence of dark matter in our universe stemming from completely different datasets. What is it? How does it binds galaxies together? How was it produced? We do not know. The dark matter paradigm lies at the interface of particle physics, astrophysics and cosmology and its fundamental nature is one of the foremost problems in science. Its interpretation in terms of elementary particles is compelling \cite{Bertone:2018krk}. That said, a wealth of searches have been conducted for signs of dark matter particles via their scattering off nuclei at underground laboratories \cite{Undagoitia:2015gya}, production at colliders \cite{Abercrombie:2015wmb} as well as in astrophysical probes for gamma-ray and cosmic-ray emissions \cite{Gaskins:2016cha}. These methods are within the three-fold dark matter searches namely, direct, indirect and collider. Nevertheless, no conclusive signal has been observed thus far, and that gave rise to several studies that revisited our common assumptions concerning the production mechanisms of dark matter and detection methods \cite{Berlin:2016vnh,Bernal:2017kxu}.    
In this work, we investigate what could be arguably seen as a new detection method, where the dark matter observable is the scattering cross section similarly to direct detection experiments, but the probe is astrophysical as occurs in indirect detection searches. In order words, it is hybrid dark matter search. 

Dark matter particles interact around once per year with detectors on Earth which makes their search rather challenging. However, extreme compact objects, such as neutron stars, arise as good targets to probe dark matter properties due to their high mass density, which are  unattainable at Earth laboratories. \cite{Gould:1989gw,Kouvaris:2007ay,Bertone:2007ae,Kouvaris:2010vv,deLavallaz:2010wp,PerezGarcia:2010ap,McCullough:2010ai,Guver:2012ba,Kouvaris:2013awa}. %The density of a neutron star is $\sim 10^{15} g cm^{-3}$ almost 3-times the normal nuclear density ($2.8\times 10^{14} g cm^{-3}$). 
The interactions between dark matter particles and neutrons can potentially increase the neutron star temperature \cite{Baryakhtar:2017dbj, Raj:2017wrv}. Thus, it is conceivable that neutron star observations can be used as a medium to infer dark matter properties and consequently probe the nature of dark matter \cite{Pshirkov:2007st,Raj:2017wrv,Ellis:2018bkr,Bramante:2017xlb,Hook:2018iia,Ellis:2018bkr,Kopp:2018jom,Bell:2018pkk}.

\begin{figure*}[h!]
\includegraphics[width=0.7\columnwidth]{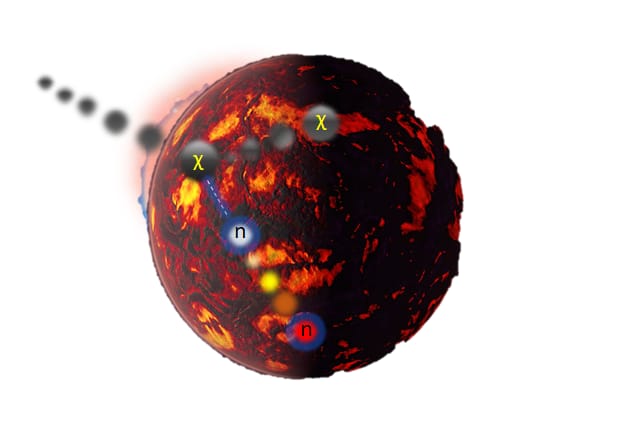}
\label{fig:xsex}
\end{figure*}

The locations of nearby neutron stars can be determined by detecting their radio pulses. The distances are estimated from pulsar dispersion measurements, estimated distances to the related supernova remnants, or observations of interstellar absorption to other stars in proximity. In three cases, parallax estimates are available \cite{Page:2004fy}. Based on the current population of neutron stars and the heat caused by dark matter-neutron interactions we have drawn in the luminosity vs age plane, the region which dark matter particles could be detected  using neutron star spectroscopy. For concreteness we show for the cases where dark matter particles interact via scalar and pseudoscalar particles that neutron stars are great laboratories to probe the nature of dark matter (see Fig.\eqref{fig:xsex} for an illustrative drawing of the process).

\section{Results}

Our reasoning relies on the simple conversion of {\bf recoil} energy to thermal energy \cite{Baryakhtar:2017dbj,Raj:2017wrv}. To do so, we have to compute the energy transferred to a neutron inside a neutron star in the dark matter-nucleon scattering process in a relativist setting. This energy corresponds to the same recoil energy inferred in direct detection experiments on Earth but taking into account relativistic effects. For this reason, neutron stars indeed constitute an orthogonal search for dark matter. In simple terms, for a typical neutron star of mass $M_\star = 10 M_\odot$ and radius $R = 10$ Km, the escape velocity (velocity in the surface of the neutron star) is around $v_{esc} = \frac{G M_\star}{R}\sim 0.1 c$, whereas the escape velocity of dark matter in our halo is around  $v_{\chi} \sim 10^{-3} c$, thus non-relativistic. Conversely, the detection of dark matter via neutron stars requires relativistic corrections. The details of the kinematic relations will be explained later on. The recoil energy in the relativistic case reads, 

\begin{equation}
\Delta E = \frac{\gamma^2m_\chi^2(v_\chi^2 + v_{esc}^2)(1-cos\theta^*)}{m_\chi^2+m_n^2+2\gamma m_\chi m_n}E_n,
\end{equation}

where $m_\chi$, $m_n$ are the dark matter and neutron masses respectively, $cos\theta$ and $E_n$ the scattering angle and the total energy of the neutron, which in the neutron rest frame is $E_n = m_n$. The final temperature acquired by the neutron star depends on the interaction rate, which is characterized by the maximum impact parameter $b_{max} = \left(\frac{2GMR}{v_\chi^2}\right)^{1/2}\left(1-\frac{2GM}{R}\right)^{1/2}$, for which the dark matter in the halo intersects a neutron star \cite{Goldman:1989nd}. The flux of dark matter passing through the neutron star is then,

\begin{equation}
\dot{m} = \pi b^2 v_\chi \rho_\chi,
\end{equation}where $\rho_\chi$ is the dark matter density in the halo. The rate at which the kinetic energy is deposited is given by,

\begin{equation}
\dot{E}=\frac{E_\chi^R \dot{m}}{m_\chi} f.
\label{Eqrate}
\end{equation}

In equation \eqref{Eqrate}, $f$ is the capture efficiency. It depends on the relation between dark matter-nucleon scattering cross section $\sigma_{\chi n}$ and a saturation cross section, $\sigma_S$, above which all the transient dark matter is captured so that,

\begin{equation}
f = min(\sigma_{\chi n}/\sigma_S, 1).
\label{s_ratio}
\end{equation}

In order words, the dark matter particle will be captured if the deposited energy exceeds its initial kinetic energy in the halo far away from the neutron star. The saturation cross-section depends on the neutron star geometric cross section $\sigma_0=\pi (m_n/M_\star)R^2$, which varies according to the dark matter mass as follows,

\begin{align}
\displaystyle
\sigma_{S} =
 \begin{cases}
	\frac{\text{GeV}}{m_\chi} \,\sigma_0
	&\text{if}~~ m_\chi < \text{GeV}, 
	\\
	\sigma_0
	&\text{if}~~\text{GeV}\leq m_\chi \leq 10^6~\!\text{GeV}, 
	\\
	\frac{m_\chi}{10^6~{\rm GeV}}\,\sigma_0
	&\text{if}~~ m_\chi > 10^6~{\rm GeV}.
 \end{cases}
 \label{eq:sigmathreshold}
\end{align}

\begin{enumerate}
	\item $\sigma_S$ for $m_\chi \!<\!\text{GeV}$, 
	the typical momentum transfer $\sqrt{2m_n \Delta E}$ is smaller than the neutron star Fermi momentum $p_F \simeq 0.45~\text{GeV} \left[\rho_{\rm NS}/(4 \times 10^{38}~\text{GeV}\,\text{cm}^{-3})\right]$. As the neutrons star is an extreme compact object composed of highly degenerate neutrons, protons and electrons the dark matter - nucleon cross section is affected by the Pauli exclusion principle reducing the number of nucleons accessible to scatter off with dark matter by a factor of $\frac{\delta p}{p_F}$ with $\delta p \sim \sqrt{\Delta E} \sim \gamma m_\chi v_{esc}$ the transferred momenta. This implies that $\sigma_S \propto  m_\chi^{-1}$.

	\item $\sigma_S$ for $\text{GeV} \!\leq\! m_\chi \!\leq\! 10^6~\!\text{GeV}$, one single scattering with depletes all the kinetic energy $\Delta E$ in the halo and gravitationally binds the dark matter particle to the neutron star. Therefore, $\sigma_S = \sigma_0$.

	\item $\sigma_S$ for $m_\chi > 10^6\!~\text{GeV}$, the kinetic energy in the halo exceeds the recoil energy given to the neutron, requiring multiple scatters to capture the dark matter. The saturation cross section is proportional to the number of scatters, 
		$\sigma_S\propto m_\chi$. 
\end{enumerate}

As aforementioned, relativistic effects should be incorporated. They will affect the apparent (redshifted) temperature of the neutron star observed on Earth \cite{Yakovlev:2004iq}, commonly known as the effective temperature of a neutron star, $T_\star$. After the thermalization, which takes less than a year, the neutron star's temperature can be described by a black-body spectrum being the thermal photon luminosity in the local reference frame of the star given by $\mathcal{L_\star}=4\pi \sigma_B R^2 T_\star^4$.  The apparent effective temperature $T_{NS}$ and luminosity $\mathcal{L_{NS}}$ as detected by a distant observer, are

\begin{equation}
T_{NS}=T_\star \sqrt{1-\frac{2GM_\star}{R}},
\end{equation}where 

\begin{equation}
 \mathcal{L_\star} = \dot{E} = \frac{E_\chi^R \dot{m}}{m_\chi} f = 4\pi \sigma_B R^2 T_{\star}^4,
\end{equation}which implies into an  apparent temperature of,

\begin{equation}
T_{NS} = \left[\frac{(\gamma^R-1)b_{max}^2v_\chi \rho_\chi}{\sigma_B R^2}\right]^{1/4}\left(1-\frac{2GM_\star}{R}\right)^{\frac{1}{2}}f^{1/4}.
\label{temperature}
\end{equation}

In equation \eqref{temperature}, $\gamma^R$ is the gamma-factor corrected by the gravitational effects as $\gamma^R  = \gamma^\infty  + \frac{GM}{R}$ with $\gamma^\infty = (1-v_\chi)^{-1/2}$. Setting the dark matter local density to be $\rho_\chi = 0.42$ GeV$cm^{-3}$, for a typical neutron star (radius and mass) that captures the entire flux of dark matter passing by we get,

\begin{equation}
T_{NS} \sim 1750f^{1/4}[K].
\end{equation}

%This low temperature matches with cooling models for neutron stars in which a thermalization process leads to temperatures of $T\lesssim 1000$ K \cite{Page:2004fy}.
As the velocity of the dark matter on the surface of the neutron star is large, the transferred momentum is much larger than the one expected on earth. The use of non-relativistic effective operators to describe the dark matter-nucleon scattering is not advised at this point. For this reason, we used simplified models where the momentum and mass of the particle that mediates the interaction between dark matter and quarks is resolved. For concreteness, we consider two different models: (i) scalar mediator; (ii) pseudoscalar mediator. We assume that the dark matter particle is a dirac fermion that interacts with quarks at equal strength, $g$, either via a scalar or pseudoscalar particle. These possibilities give rise to three possible lagrangians and results that we will discuss in the next section. In a nutshell, the idea of probing dark matter via neutron stars is summarized as follows: 

\begin{itemize}
    \item (i) The thermal emission of neutron stars have not been directly measured thus far (apart from a dozen cases \cite{Page:2013hxa}). Their (non-thermal) luminosity can be inferred from X-ray flux when distance is known (bolometric luminosity), estimated from breakage or through models;
    \item (ii) Future telescopes  with  sensitivity to infrared radiation with wavelength of $\lambda \sim 3 \mu m$ may eventually directly measure neutron star's temperature down to $1000 $~K.;
    \item (iii) Dark matter models can heat up neutron stars up to $\sim 1750$~K.;
    \item (iv) By comparing the theoretical prediction and future measurements of neutron star temperature one can probe the presence of dark matter interactions, and consequently its nature. 
\end{itemize}

\begin{figure*}[h!]
\includegraphics[width=0.49\columnwidth]{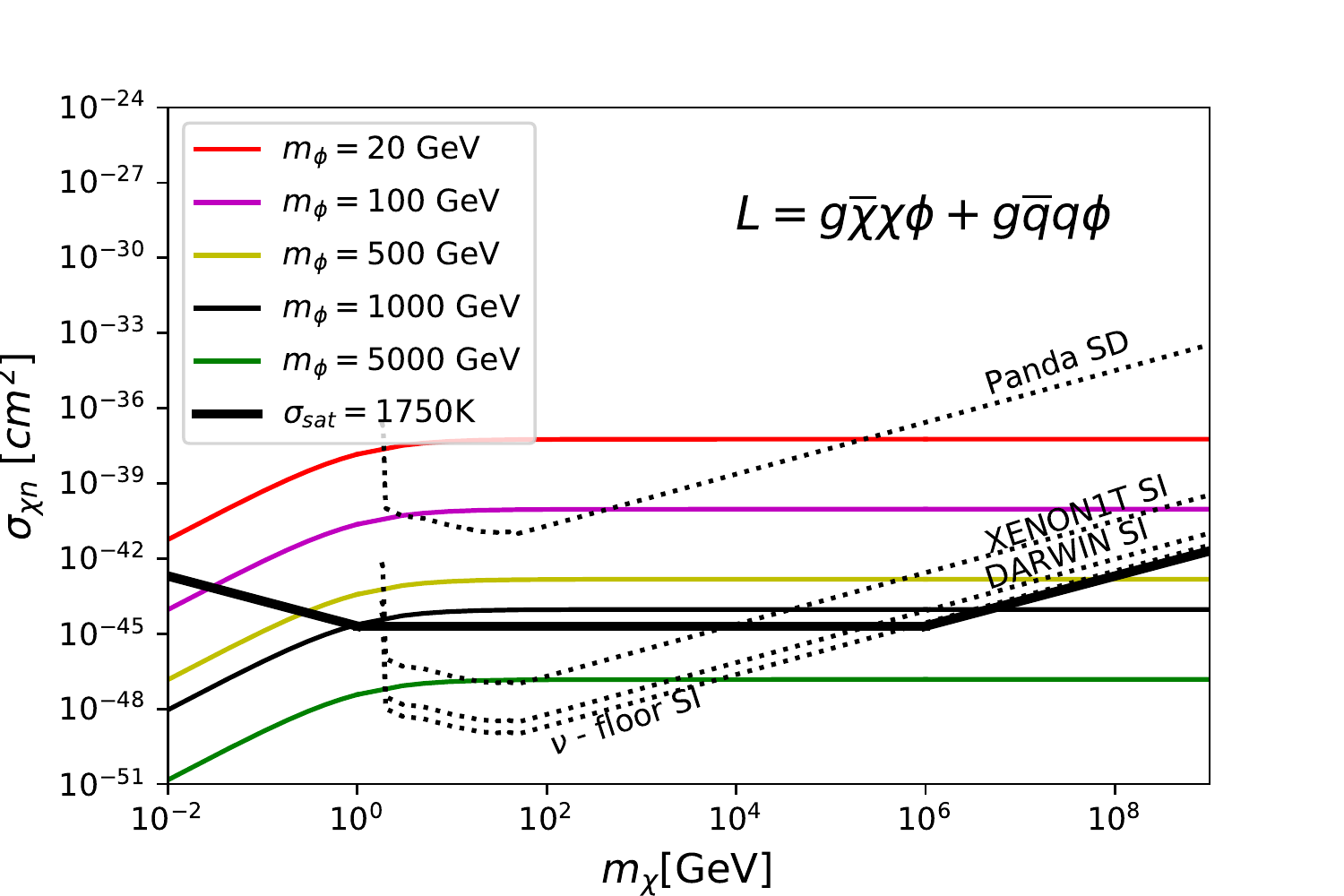}
\includegraphics[width=0.49\columnwidth]{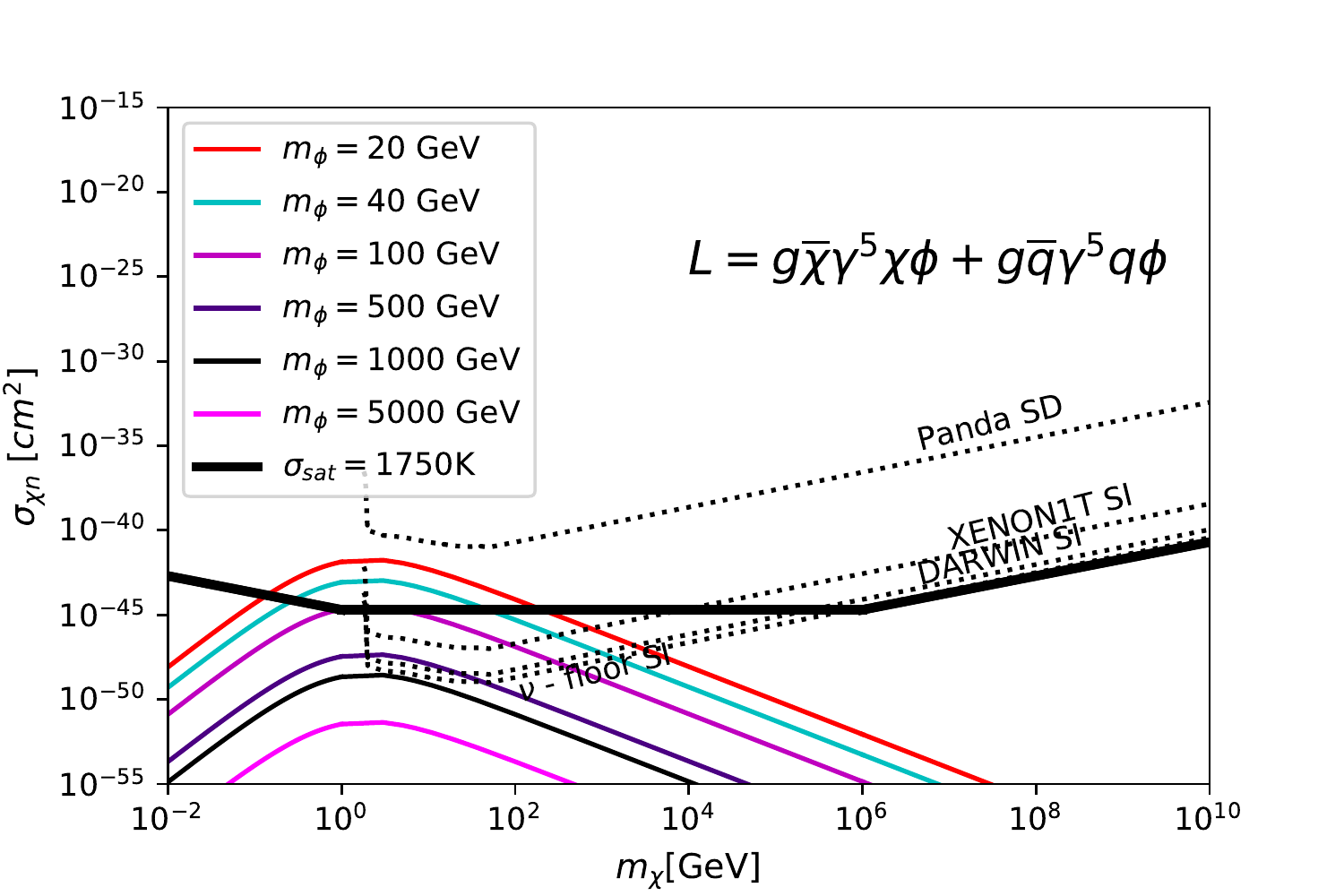}
\includegraphics[width=0.5\columnwidth]{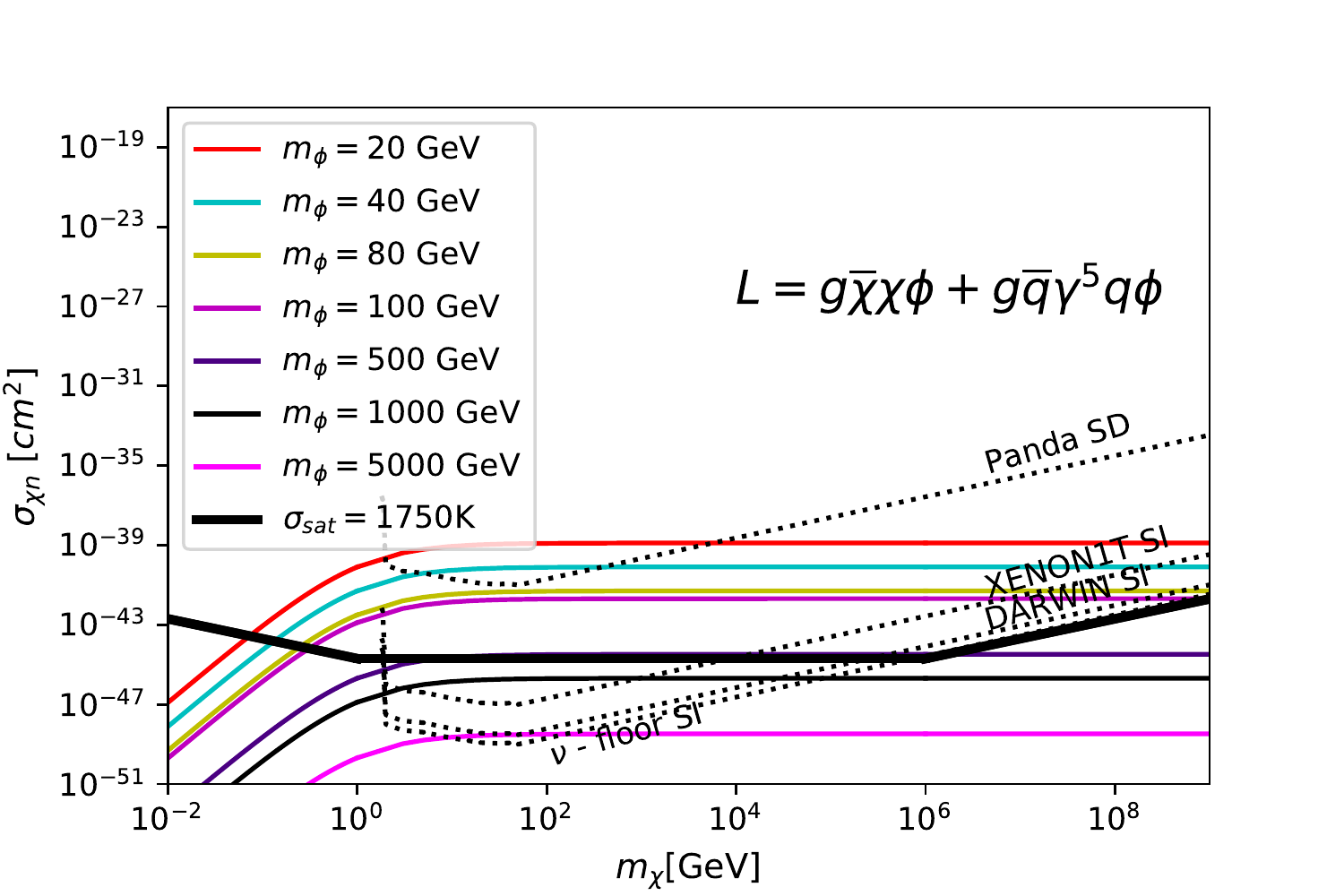}
\caption{Dark matter-neutron cross-section $\sigma_{\chi n}$ as function of the dark matter mass for several values of the scalar and pseudo-scalar mediators. The solid lines correspond to different scenarios while pointed lines correspond to upper limits on $\sigma_{\chi n}$ imposed by PANDA, LUX and XENON experimental collaborations. The solid blue line $\sigma_{sat}$ corresponds to the saturation cross-section for which all the dark matter particles passing thorough the neutron star are captured then providing the maximum heating.}
\label{fig:xsex}
\end{figure*}

\begin{figure}[h!]
\includegraphics[width=0.49\columnwidth]{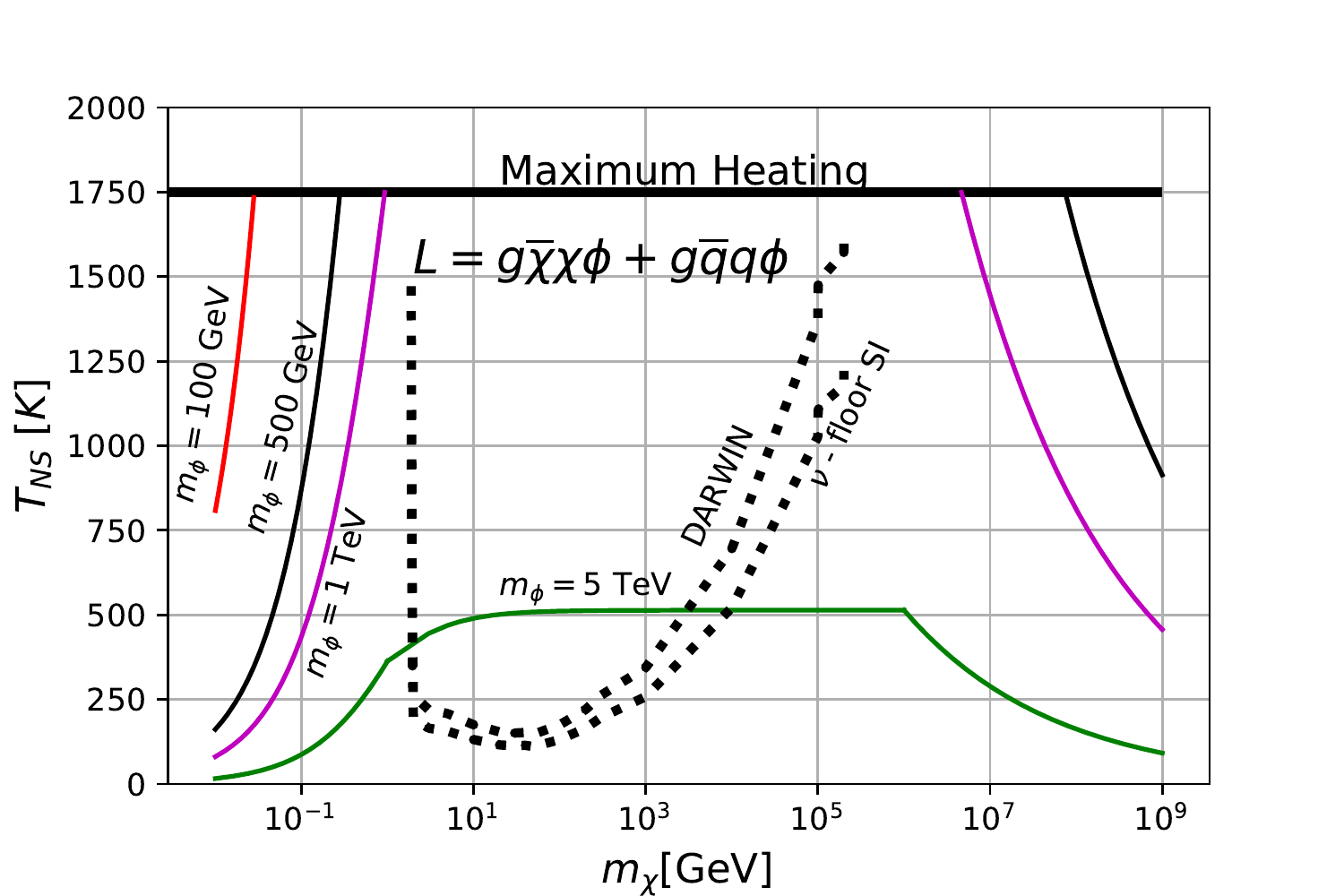}
\includegraphics[width=0.49\columnwidth]{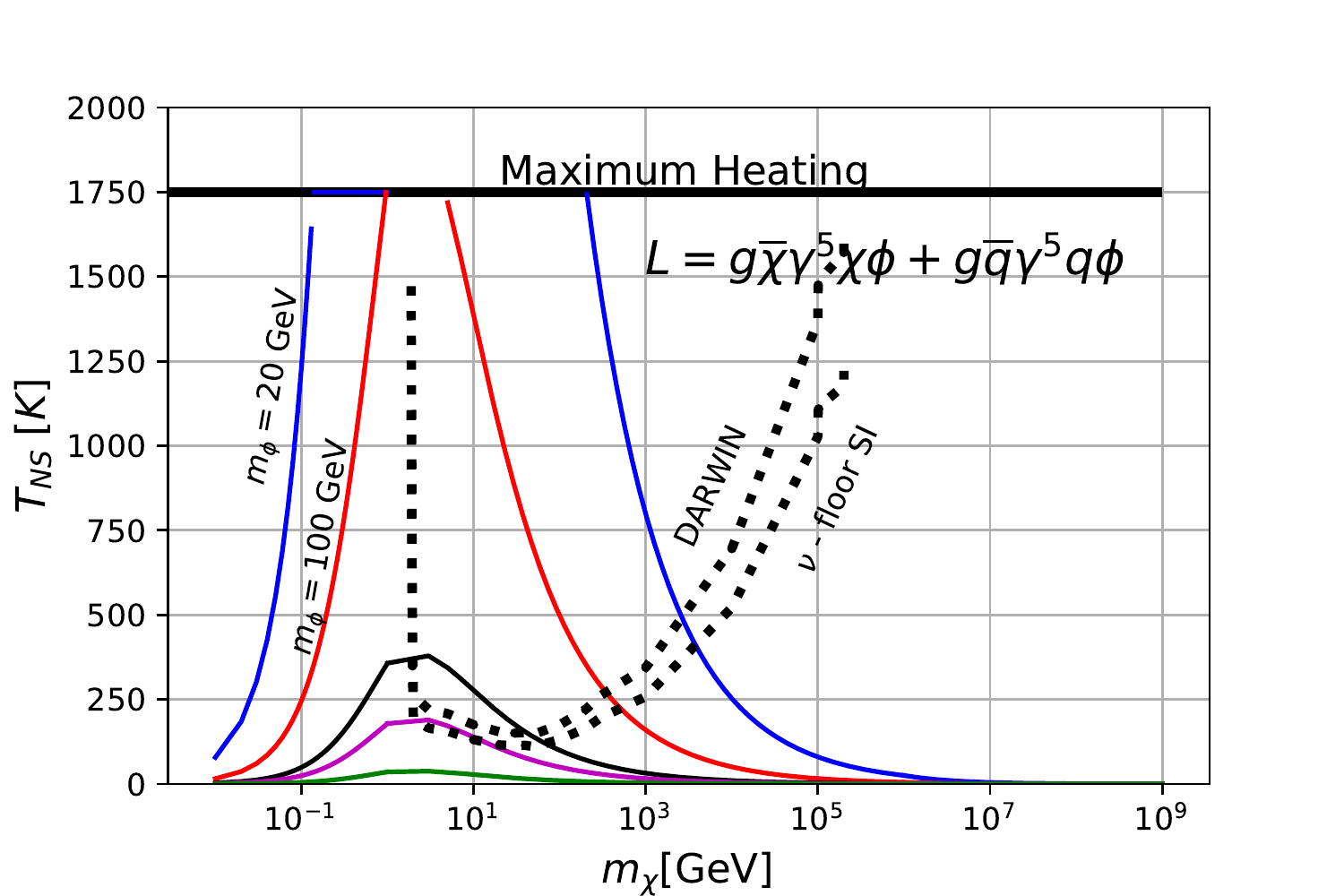}
\includegraphics[width=0.5\columnwidth]{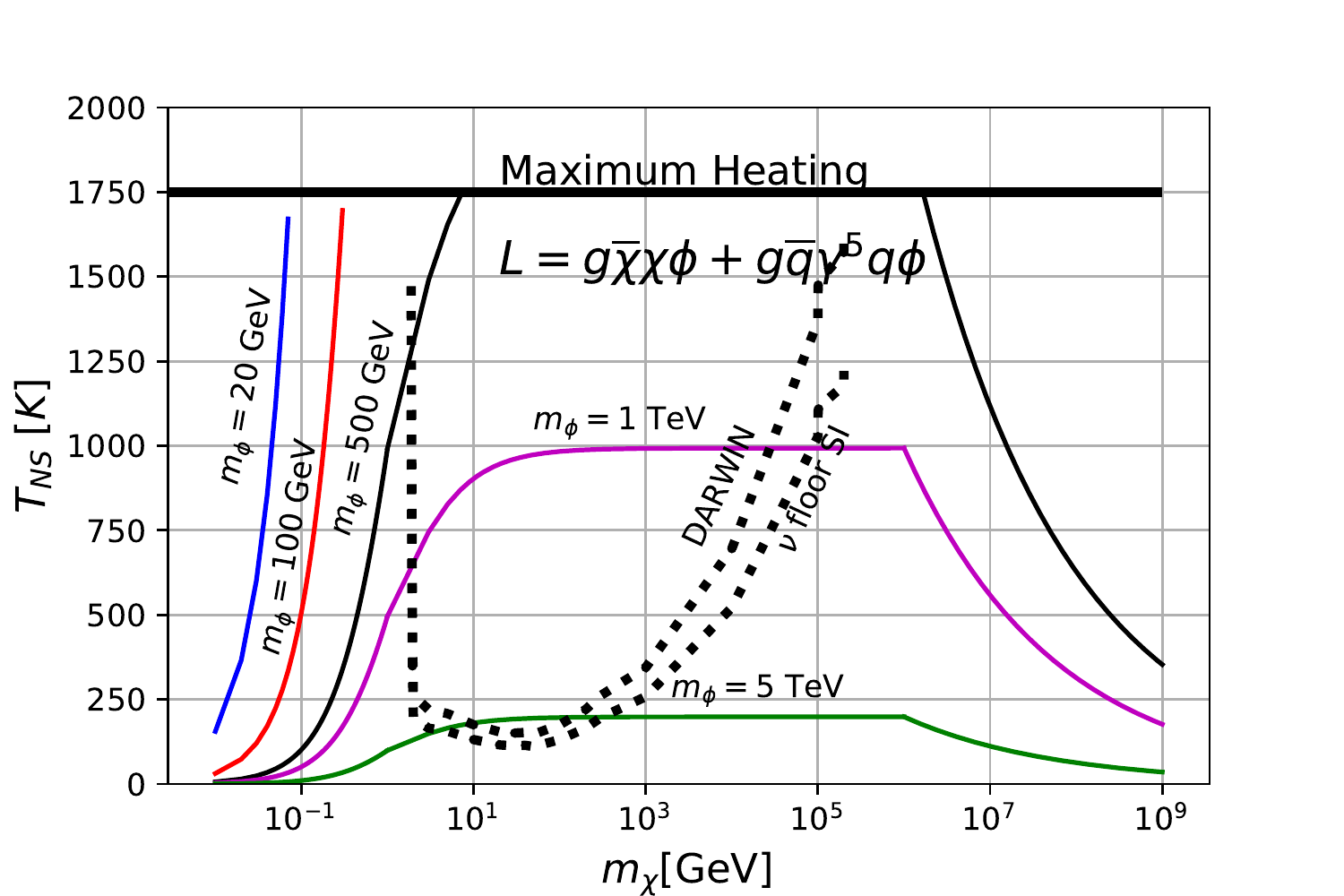}
\caption{Neutron star apparent temperature due to the dark kinetic heating as function of the dark matter mass for several values of the scalar and pseudo-scalar mediators. Solid lines correspond to the neutron star black-body temperature while the pointed line correspond to the upper limit on spin-independent $\sigma_{\chi n}$ translated to temperature of the XENON experimental collaboration.}
\label{fig:TEM}
\end{figure}
\section{Discussion} 

In order to grasp the real importance of neutron star observations to dark matter we plotted in FIG.~\ref{fig:xsex} the dark matter-neutron scattering cross section as a function of the dark matter mass for three simplified dark matter models that encompasses dark matter interactions with quarks via either scalar or pseudoscalar mediators. We will start our reasoning with model which neutron stars are less sensitive to highlight its impressive relevance. \\

The black dotted lines represent the most stringent current limits and projected sensitivities from direct detection experiments on the dark matter-nucleon scattering, which can be spin-independent or spin-dependent. We also exhibited the neutrino floor for a XENON target with a dashed black line. The thick black solid line delimits the saturation cross section which yields the maximum heating. The color lines account for the dark matter-neutron scattering for different values of the mediator mass that ranges from $20$~GeV to $5000$~GeV.\\

We emphasize that color lines above the {\it thick black solid line} which delimits the saturation cross section that heats up the neutron star to a temperature of $1750$~K, whereas the lines below this {\it thick black solid line} produce a heat below $1750$~K. The precise amount of heat produced by in the parameter space below the solid black line varies as we will later on. \\

The first model represented in the first panel of FIG.~\ref{fig:xsex}, which is for a scalar mediator, can produce a sizable heating of the neutron star up to $1750$K for: (i) $m_\chi < 1$ GeV and $m_\phi < 1$ TeV;(ii) $m_\chi > 10^7$ GeV and $m_\phi \sim 1$ TeV. If future observations can measure neutron stars temperatures down to roughly 1000K, we will be able to determine the presence of dark matter interactions by finding a plateau in neutron star temperature distribution at $\sim 10^3$ K. It is remarkable that for $m_{\chi}<1$~GeV neutron stars will constitute the best probe for dark matter. A similar conclusion we can drawn for $m_{\chi}> 10^6$~GeV. It is well-known that direct detection experiments will struggle to push their sensitivity below the neutrino floor \cite{Dent:2016iht,Ng:2017aur,Boehm:2018sux,Papoulias:2018uzy}, but with neutron star spectroscopy one can easily overcome this issue. Already for the scalar mediator case, we conclude that neutron stars can be seen as potential dark matter detectors capable of probing dark matter beyond the neutrino floor. \\

We have performed a similar analysis for the case of pseudoscalar mediator and mixing of scalar and pseudoscalar interactions. These are eye-catching examples of how powerful neutron star observations can be. Hopefully, with an aggressive progress we will be able to measure neutron stars temperature down to 1000K to be able to probe dark matter particles with an impressive sensitivity.  \\

To better understand the relevance of the lines shown in FIG.~\ref{fig:xsex} below the saturation cross section curves, we displayed in FIG.\ref{fig:TEM} the evolution of the neutron star temperature as a function of the dark matter and mediator masses. Again, we will focus on the most conservative setup, which neutron stars are the least sensitive to. Looking at the green line in the first panel of FIG.~\ref{fig:xsex} which is for $m_{\phi}= 5000$~GeV, we notice that direct detection experiments will not be able to probe the dark matter model for  $m_{\chi} < 1$~GeV and $m_{\chi} > 10^4$~GeV. Now by glancing at FIG.~\ref{fig:TEM} we notice that the region with $m_{\chi} < 1$~GeV yields $T_{NS} < 500$~K, whereas the one with $m_{\chi} > 10^4$~ GeV produces $T_{NS} \sim 500$~K. As mentioned earlier, direct detection experiments will have no sensitivity to this region of parameter space because they wind up in a very small scattering cross section, below the neutrino floor. Albeit, future neutron observations stand as a hope to eventually probe such dark matter models. We underscore that such dark matter models with masses above few TeVs have become object of intensive search at indirect detection experiments \cite{Berlin:2016vnh,Berlin:2016gtr,Profumo:2016idl}. If one does the same exercise for the other two models, the importance of neutron stars observations will be further strengthened. It is undeniable that future measurements of neutron stars temperature will represent a new dark matter detection method.   \\

In what follows we will give an astrophysical introduction to the physics of neutron stars and later detail our calculations.

\section{Neutron star temperature}

Neutron stars arise from the collapse of few solar mass stars, reaching at their birth 
temperature as large as $\sim 10^{11} \rm{K}$ \cite{Page:2004fy}.
Yet their matter is so dense that the degeneracy pressure is larger than the
temperature in all of the star but for a small fraction in the outer layer.
Heat is produced in the core and dissipated into space through the atmosphere, which
imprints its temperature into the radiation, and the cooling
process occurs rapidly first in a neutrino driven phase (for a duration $\sim 10^4\,\rm{years}$)
then in a photon driven one at later times ($>10^5\,\rm{years}$) \cite{Potekhin:2015qsa}.
Studying the evolution of temperature with time is complicated and depends on neutron star equation
of state, chemical composition, mass, magnetic field but some qualitative general trends can be summarized.
As discussed in \cite{Page:2013hxa,Ofengeim:2017cum}at the onset of photon-dominated  cooling process, the temperature $T$ drops exponentially
with time (or at most with 
a power low $T\propto t^{-1/\alpha}$, with $\alpha\ll 1$
\cite{Page:2004fy}) hence it is likely that in absence of dark matter
heating of neutron stars their temperature may drop below the value of a thousand Kelvin.\\

In the collapse process the resulting neutron star acquires a fast rotation, with period in
the range $10^{-3}$ to few seconds, and a magnetic dipole. If the magnetic
dipole has a misalignment angle $\alpha$ with the rotation axis, the rotating
neutron star will emit an energy rate $\dot E$ given by
\be
\label{eq:edotBNS}
\dot E=\frac 23 B^2R_{NS}^6\Omega^4\sin^2\alpha\,,
\ee
being $B$ the value of the magnetic field, $R_{NS}$ the neutron star radius and $\Omega=2\pi/P$
its angular velocity for period $P$ and $\alpha$ the angle between the magnetic and rotational axis.
By equating the emission energy rate to the loss of rotational energy $\dot E_{rot}=-I\Omega\dot\Omega$
($I$ being the moment of inertia)
one gets an estimate of the equivalent magnetic field
$B_{eq}\simeq 3.2\times 10^{19}(P\dot P)^{1/2}\rm{G}$, assuming
$\sqrt{I_{35}}/(R_6^3\sin\alpha)=1$, with $I_{35}\equiv I/(10^{35}\rm{gr\cdot km^2})$ and $R_6\equiv R_{NS}/(10\,\rm{km})$.\\

For neutron stars seen as pulsar the rotation period $P$
and its time derivative $\dot P$ are well measured, see e.g. the online catalog
%http://www.atnf.csiro.au/research/pulsar/psrcat, 
\cite{Manchester:2004bp},
enabling a estimate of their age $\tau_c$
\be
\label{eq:tauc}
\tau_c\equiv P/(2\dot P)\,,
\ee
which is defined as the \emph{characteristic age}.
Still within the assumption that the braking is entirely due to the electromagnetic emission one can study the derived
neutron star emission rates versus their age to extrapolate at what age they will reach an emission
compatible with a thermal radiation at $T\sim 10^3\,{\rm K}$.\\

Emission spectra of neutron star are usually complicated and non-thermal, involving detailed structure of their
atmosphere, however they may contain a thermal component that can dominate in a frequency band and which can then be isolated
and measured.
Thermal emission from isolated neutron stars have been first detected by X ray telescopes HEAO2-\emph{Einstein} and EXOSAT \cite{Cheng:1983,Brinkmann:1987},
in the last decades of the twentieth century X-rays have been effectively used to pinpoint neutron star emission,
with ROSAT and ASCA \cite{ogelman95}, and the in the optical UV-range with the \emph{Hubble Space Telescope}.
Later Chandra \cite{Weisskopf:2002sy,Pavlov:2002ec} and XMM-Newton \cite{Becker:2002wf}
allowed the study of neutron stars with thermal emission in X-rays, see \cite{Potekhin:2014fla} for a review.\\

\begin{figure}
\includegraphics[width=.8\linewidth]{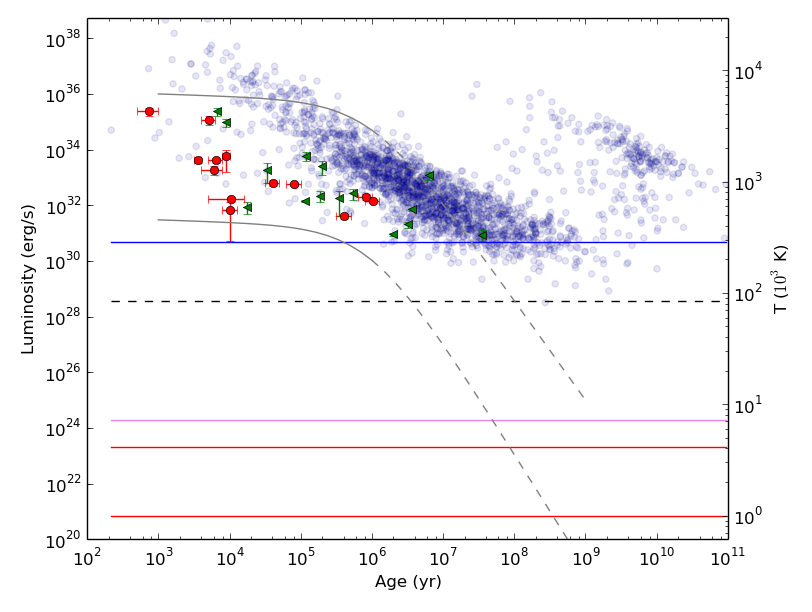}
\caption{Blue semi-transparent dots are energy emission rate corresponding to the
 observed neutron star braking vs. $\tau_c$ defined in eq.~(\ref{eq:tauc}).
It is visible the presence of two populations of neutron stars, one with larger emission rate for larger ages are
recycled pulsars, see text for explanations. 
Red circles with error bars represent bolometric luminosity versus kinetic estimate of neutron star age.
Green triangles represent bolometric luminosity versus $\tau_c$, which can be
considered as an upper limit to the actual neutron star age.
Horizontal solid lines, from top to bottom, represent
respectively the energy rate for an equivalent thermal emission with $T\simeq 3\times 10^5{\rm Kelvin}$, corresponding to the maximum
wavelength in the $X$ spectrum ($\lambda =10^{-6}{\rm cm}$), the second line from the top $T\simeq 7\times 10^3{\rm Kelvin}$
($\lambda=400 {\rm nm}$) corresponding to the UV $\to$ visible transition, then the line at $T\simeq 4\times 10^3{\rm Kelvin}$ ($\lambda= 700{\rm nm}$, visible 
$\to$ IR) and finally the line corresponding to the thermal emission at $T=10^3{\rm Kelvin}$ ($\lambda\simeq 3{\rm \mu m}$). The dashed horizontal line corresponding to luminosity $\simeq 3.5\times 10^{28}$ erg/sec denotes the onset of
the \emph{graveyard} region for pulsars.
The grey solid lines delimit the region of thermal emission vs. age
according to the model in \cite{Vigano:2013lea}, which turn dashed in the
region we extrapolated at low luminosity, where an
exponential luminosity decay with time $L\propto \exp(-t/\tau)$ (power law $L\sim t^{-4}$)
for the lower (upper) dashed grey line
has been used for definiteness.}
\label{fig:cool}
\end{figure}

In fig. \ref{fig:cool} we give an overview of the status of present knowledge of pulsar observation and emission.
The semi-transparent blue circles show the emission rate assuming that its completely accounted
by the braking of the pulsars versus the age $\tau_c$ defined above.
The horizontal solid lines represent temperatures for thermal emission corresponding to different wavelength, ranging from
$X$ rays to the infrared.\\

Note the presence in the luminosity versus age plane of two distinct population of neutron stars, the old ones with high
luminosity corresponding to \emph{recycled pulsars} which have been span-up due to matter accretion from a companion.\\

Neutron star luminosity can be better estimated when the bolometric luminosity can be directly measured, as it
happens in few cases taken from \cite{Vigano:2013lea},
and also its age can have a better estimate than $\tau_c$ when a kinematic
determination is possible: observing a neutron star receding from a galactic center and measuring its velocity is a direct
way to compute its age, when such observation are possible.
The kinematic age determination lead consistently to lower values than $\tau_c$, implying that the magnetic field also
decays in time, causing more emission at early stages.\\

Borrowing the result of the simulation reported in \cite{Vigano:2013lea} (where predicted thermal emission rates are shown up to age $\sim 10^6$ years) and
extrapolating cooling models to infer the thermal emission rate
(hence the temperature) for older stars, it is plausible that temperature of
$10^3$ K or less can be reached due to the rapid fall of temperature at late times.\\

Note that pulsar observation stops at a luminosity value
$\dot E_{gy}\simeq 3.5\times 10^{28}{\rm erg/sec}$, which is interpreted 
in the standard scenario as the condition for the magnetic field and angular velocity
reaching the critical values at roughly constant $B/P^2$ at which the pulsar
emission mechanism stops and neutron stars enter the \emph{graveyard} region.
After that the spin down of the neutron star should be negligible but the magnetic
field should continue to decrease.
A temperature of $\sim 10^3$ K represents a very small value compared to presently detected neutron stars temperatures which
are in the range of $\sim 10^7$ K, since they have been observed via their UV and X-ray emission.\\

To detect thermal radiation at $T\sim 10^3$ K one needs a telescope with sensitivity to infrared radiation with wavelength $\lambda \simeq 3\mu$m.
In this window fluxes as low as $\sim {\rm n Jy}$ \footnote{${\rm Jy}=10^{-23} {\rm erg/(sec\cdot cm^2\cdot Hz)}$}
can be detected by observatories as the James Webb space telescope
\cite{Kalirai:2018qfg}.
For comparison the thermal flux ${\cal F}_T$ sourced by a neutron star at
distance $D$ from the observer is
\be
{\cal F}_T=\sigma_B \pa{\frac{R_{NS}}{D}}^2 T^4\simeq 6\times
10^{-24}{\rm \frac{erg}{cm^2\,sec}}\pa{\frac T{10^3\,{\rm K}}}^4
\pa{\frac{R_{NS}}{10{\rm km}}}^2
\pa{\frac D{1{\rm kpc}}}^{-2}\,,
\ee
and considering a detector bandwidth of the same order of the frequency $f\simeq 10^{14} {\rm Hz}\pa{T/10^3{\rm K}}$, we are
still few order of magnitude short in sensitivity for galactic sources.

\section{Methods}

\subsection{Interactions}

We have considered a set of simplified models including scalar and pseudo-scalar mediators. We are interested in exploring new constraints or proves of very light and very heavy mediators, then simplified scenarios are a good starting point. In equations \eqref{scalar}, \eqref{mixscalar} and \eqref{psudoscalar} there are three different simplified models for a fermionic dark matter $\chi$ and scalar and pseudo-scalar mediators $\phi$ being the equation \eqref{mixscalar} a mixture.  

\begin{equation}
 L \supset g \Bar{\chi}\chi\phi + g \Bar{q}q\phi  
\label{scalar} 
\end{equation}

\begin{equation}
 L \supset g \Bar{\chi}\chi\phi + g \Bar{q}\gamma^5 q\phi  
\label{mixscalar} 
\end{equation}

\begin{equation}
 L \supset g \Bar{\chi}\gamma^5\chi\phi + g \Bar{q}\gamma^5 q\phi  
\label{psudoscalar} 
\end{equation}where $g$ is the coupling constant. We will take as a benchmark point $g=1$ where other scenarios can be re-scaled straightforwardly.  \\

In equations \eqref{s_scalar}, \eqref{s_mix} and \eqref{s_psscalar} there are the scattering dark matter-neutron cross-sections for the three cases mentioned above,

\begin{equation}
    \sigma_{S} = \frac{g^4}{\pi}\frac{\mu_{\chi n}^2}{(p_{\chi n}^2-m_\phi^2)^2},
\label{s_scalar}
\end{equation}

\begin{equation}
    \sigma_{mix} = \frac{g^4 p_{\chi n}^2}{16\pi m_n^2}\frac{\mu_{\chi n}^2}{(p_{\chi n}^2-m_\phi^2)^2},
    \label{s_mix}
\end{equation}

\begin{equation}
    \sigma_{PS} = \frac{g^4 p_{\chi n}^4}{64\pi m_\chi^2 m_n^2}\frac{\mu_{\chi n}^2}{(p_{\chi n}^2-m_\phi^2)^2}.
\label{s_psscalar}
\end{equation}

Notice that each scenario has a different $m_\chi$ dependence, and this will lead to a different neutron star temperature via equations \eqref{s_ratio} and \eqref{temperature}.  \\

We have used then the relation between temperature and the dark matter-neutron cross-section, $\sigma_{\chi n}$, to investigate where our models are excluded or not by future observations of neutron stars as well as to compare them with the current constraints on $m_\chi$ imposed by experiments of direct detection on earth. 

\subsection{Energy Deposition}

\begin{figure}[h!]
\includegraphics[width=\columnwidth]{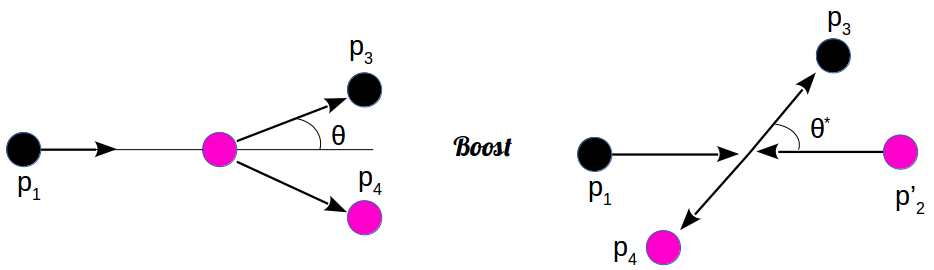}
\caption{Schematic diagram of the dark matter-neutron collision in the neutron star frame (left) and in the center of mass frame (right).}
\label{fig:LEP}
\end{figure}

For a 4-vector of a single particle of rest mass $m$ and velocity $v$ we have,

\begin{equation}
P^\mu = (E,\vec{p}) \Longrightarrow P_\mu^2 = m^2 = E^2 - |\vec{p}|^2,
\end{equation}

then

\begin{equation}
E^2 = m^2- |\vec{p}|^2 = m^2 - m^2 v^2 = m^2 (1-v^2) = \gamma^2 m^2 .
\end{equation}

It is convenient to define the boost factors in terms of energy and momentum once the velocity of the light is taken to be equal one where,

\begin{equation}
\gamma = E / m = \frac{E}{\sqrt{P_\mu P^\mu}}.
\end{equation}

On the other hand,

\begin{equation}
\beta = \frac{v}{c}=\frac{\gamma mv}{\gamma mc}=\frac{\gamma mvc}{\gamma mc^2}=\frac{pc}{E}=\frac{p}{E}
\end{equation}

The dark matter and the nucleon have the initial 4-momentum in the nucleon rest-frame,

\begin{equation}
P_1^\mu = (E_1,\vec{p_1}), \hspace{10mm} P_2^\mu = (E_2,\vec{0}) .
\end{equation}

The total 4-momentum of the system reads,

\begin{equation}
\begin{split}
P_T^\mu = (E_1 + m_2, \vec{p_1}) \Longrightarrow  P_T^\mu P_T\mu & =  (E_1+m_2)-|\vec{p_1}|^2 \\ = m_1^2+m_2^2+2E_1m_2.\\
\end{split}
\end{equation}

Therefore, the $\gamma$ and $\beta$ factors read,

\begin{equation}
\gamma = \frac{E_T}{\sqrt{P_T\mu P_T^\mu}}= \frac{E_1+m_2}{\sqrt{m_1^2+m_2^2+2E_1m_2}},
\end{equation}and,

\begin{equation}
\beta = \frac{p_T}{E_T} = \frac{\gamma m_1v_1}{E_1+m_2}.
\end{equation}

\begin{figure}[t!]
\includegraphics[width=\columnwidth]{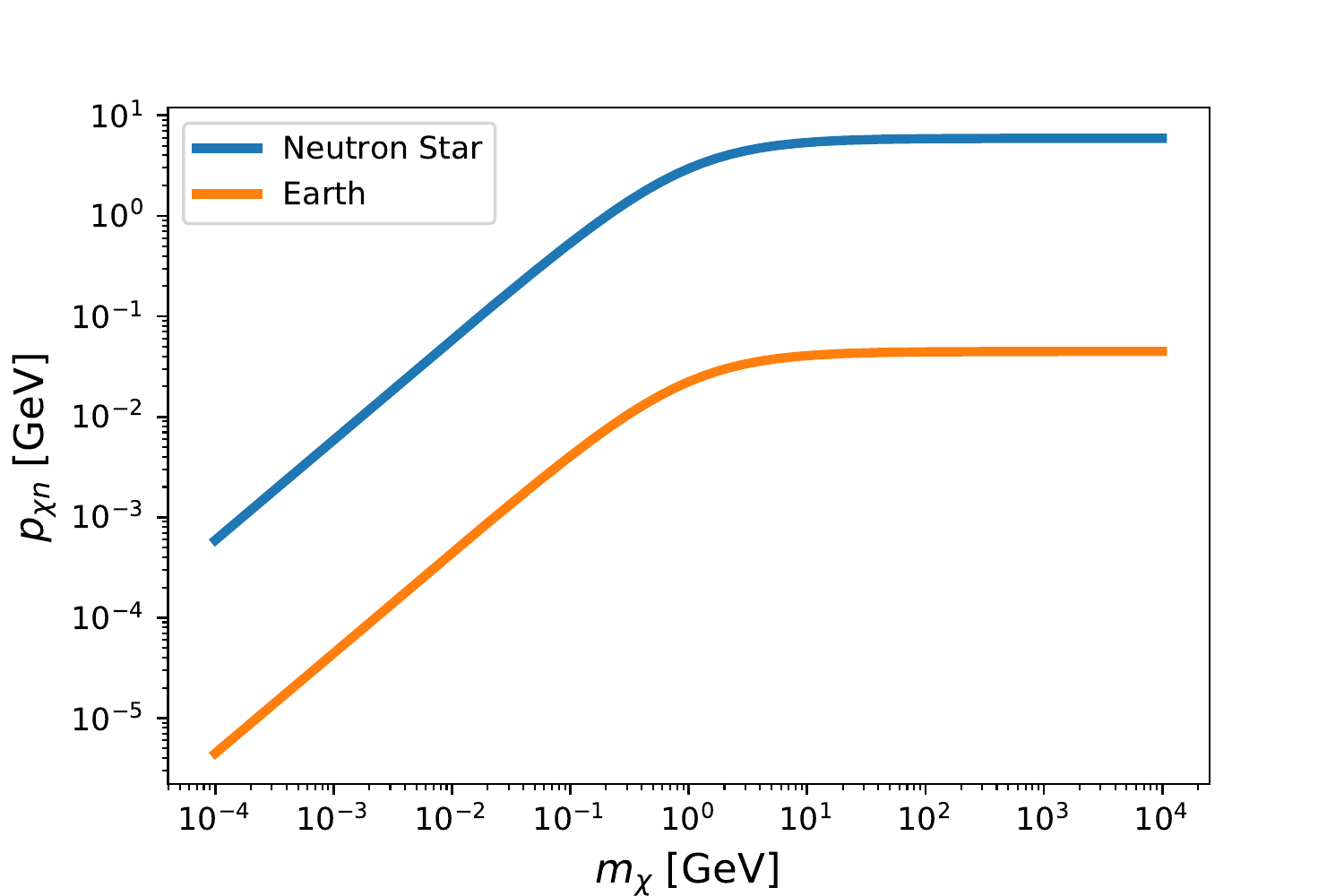}
\caption{In the figure we show the momenta that is transferred to the neutron by the scattering with the dark matter particle $p_{\chi n} = \sqrt{2m_n \Delta E}$ where $m_n$ is the nucleon mass typically 1 GeV. From it, we can see that the typical momenta rise $p_{\chi n} \sim 10$ GeV for masses $m_\chi > 20$ GeV which is two order of magnitude larger than the one expected on earth $p_{\chi n} \sim 10$ KeV.}
\label{fig:momenta}
\end{figure}

It is convenient to work in the center-of-mass frame (CoM) and then use the boost factors to pass from one frame to the other. Let the prime quantities be the CoM ones, then the 4-momentum of the system in the CoM reads,

\begin{equation}
P_2^\mu = (E_2^\prime,\vec{p_2}^\prime) 
\end{equation}

The boost of Lorentz yields,

\begin{equation}
E_2^\prime = \gamma (E_2-\beta |\vec{p_2}|) = \gamma E_2, \hspace{2mm} |\vec{p_2}^\prime| = \gamma (|\vec{p_2}|-\beta E_2)=-\gamma\beta E_2,
\end{equation}

\begin{equation}
\begin{split}
E_4 = \gamma(E_2^\prime + \beta p_2^\prime cos\theta^*) &=  \gamma(E_2^\prime + \beta (-\gamma\beta E_2) cos\theta^*) \\ =\gamma^2(1-\beta^2 cos\theta^*)E_2,
\end{split}
\end{equation}

\begin{equation}
E_4 = \frac{(E_1+m_2)^2}{m_1^2+m_2^2+2E_1m_2}\left[1-\frac{(\gamma m_1v_1)^2}{(E_1+m_2)^2}cos\theta^*\right],
\end{equation}

\begin{equation}
E_4 = \frac{E_1^2+m_2^2+2E_1m_2-(\gamma m_1v_1)^2cos\theta^*}{m_1^2+m_2^2+2E_1m_2}E_2,
\end{equation}

then with the total energy $E_1 = \gamma m_1$ and using the fact that $E_1^2-m_1^2 = |\vec{p_1}|^2 = (\gamma m_1 v_1)^2$ (relativistic momentum) we can obtain the energy that the dark matter transfer to the nucleon as the difference of the DM-energies before and after the collision $\Delta E_\chi$,

\begin{equation}
\Delta E_\chi = E_4 - E_2 ,
\end{equation}

\begin{equation}
\Delta E = \frac{\gamma^2m_1^2v_1^2(1-cos\theta^*)}{m_1^2+m_2^2+2\gamma m_1 m_2}E_2.
\end{equation}

The total kinetic energy that can be deposited by dark matter is approximately the kinetic energy at the surface of the neutron star,

\begin{equation}
E_\chi = m_\chi + K_\chi =\gamma m_\chi \longrightarrow K_\chi = m_\chi (\gamma -1)
\end{equation}

The energy of the incident dark matter particle changes due to the gravitational interactions. We can estimate an effective gamma factor at the surface of the neutron star by using classical energy conservation,

\begin{equation}
K_\chi^\infty = k_\chi^R - \frac{GMm_\chi}{R},
\end{equation}

\begin{equation}
m_\chi (\gamma^\infty -1) = m_\chi (\gamma^R -1) - \frac{GMm_\chi}{R},
\end{equation}which implies into,

\begin{equation}
\gamma^R  = \gamma^\infty  + \frac{GM}{R},
\end{equation}where for a typical neutron star $\gamma^R = 1.35$. The energy that a typical DM particle has at the surface of a neutron star is,

\begin{equation}
E_\chi^R=\gamma^R m_\chi, 
\end{equation}which is results into,

\begin{equation}
E_\chi^R= \left(\gamma^\infty  + \frac{GM}{R}\right) m_\chi.
\end{equation}

Hence, the velocity of the incoming dark matter is in terms of the escape velocity ($v_{esc}$) and taking into account the gravitational interactions,

\begin{equation}
v_\chi^2= v_\infty^2  + \frac{2GM}{R}  = v_\infty^2 + v_{esc}^2,
\end{equation}where the typical recoil energy of the scattered neutron change to be,

\begin{equation}
\Delta E = \frac{\gamma^2m_1^2(v_\infty^2 + v_{esc}^2)(1-cos\theta^*)}{m_1^2+m_2^2+2\gamma m_1 m_2}E_2.
\end{equation}

This typical recoil energy was used in the derivation of our numerical results. We highlight that our phenomenology and numerical finding have an orthogonal approach to others investigated in the literature \cite{Ciarcelluti:2010ji,Garani:2018kkd,Choi:2018axi}.

\section{Acknowledgements} 
This work was supported by MEC, UFRN and ICTP-SAIFR FAPESP grant 2016/01343-7. We thank the High Performance Computing Center (NPAD) at UFRN for providing computational resources and to Sergio Camargo for the drawing of the neutron star.

\section{Authors contributions} FSQ and DC conceived the study. DC did all the phenomenology. RS performed the link to the current neutron star catalog. All authors contributed to the writing of the manuscript. 

\bibliographystyle{unsrtnat}

%\begin{thebibliography}{99}

\bibliography{main}

%\end{thebibliography}

\end{document}